# Universal Generator for Complementary Pairs of Sequences Based on Boolean Functions

S.Z. Budišin

IMTEL, Belgrade, 11070 Serbia

P. Spasojević

WINLAB, Electrical and Computer Engineering Department, Rutgers University, NJ 08854 USA

## Abstract

We present a general algorithm for generating arbitrary standard complementary pairs of sequences (including binary, polyphase, M-PSK and QAM) of length $2^N$ using Boolean functions. The algorithm follows our earlier paraunitary algorithm, but does not require matrix multiplications. The algorithm can be easily and efficiently implemented in hardware. As a special case, it reduces to the non-recursive (direct) algorithm for generating binary sequences given by Golay, to the algorithm for generating M-PSK sequences given by Davis and Jedwab (and later improved by Paterson) and to all published algorithms for generating QAM sequences. However the algorithm does not solve the problem of sequence uniqueness (except for the trivial M-PSK case), which must be treated separately for each QAM constellation.

## Introduction

Golay introduced binary complementary sequences in [10]. They were generalized to complementary sets in [16] and polyphase sequences in [9,15]. More recently, QAM complementary sequences were studied in [6,7,11,14]. We distinguish two different families of algorithms for generation of complementary sequences. The non-recursive (direct) algorithms where first introduced by Golay [10]. Davis and Jedwab [8] gave the corresponding algorithm for M-PSK sequences (later improved by Paterson [13]) and showed that complementary sequences are closely related to Reed-Muller codes. This approach was later followed by all researchers in the field of QAM complementary sequences [6,7,11,14]. Golay also introduced an alternative family of algorithms. These are recursive algorithms that generate longer sequences from shorter ones. The first recursive algorithm for generating standard polyphase sequences (including M-PSK) was introduced by [1] and was soon followed by a recursive algorithm for generating multilevel sequences [2], which is related to QAM algorithms that appeared much later. These early recursive algorithms were all formulated in the time domain. An important application of these algorithms was the efficient correlator [3]. Recently, recursive algorithms where reformulated in the Z transform domain [12]. Matrix representation of complementary sequences was first used [16]. In [4], a compact formulation using a matrix of Z transforms led to an observation that complementary sequence pairs (and sets) can be represented via paraunitary (PU) matrices. This work resulted in a general algorithm for generating QAM sequences for regular constellations including rectangular and hexagonal ones [5]. A corresponding efficient correlator was easily derived for all these sequences. However, the PU-based implementation of the sequence generator is not as efficient as the direct algorithms based on Boolean functions. Here, we introduce a nonrecursive (direct) method for generation QAM complementary sequences based on Boolean functions. This method is closely related to the PU-based algorithms from [4].

## Definitions and Notation

We use the widely accepted definitions of sequences, autocorrelation functions and complementarity. A sequence of length $L$ is denoted by $a(n)$ for $n=0…L-1$. Its autocorrelation function is denoted by $\mathcal{R}(a(n)) = R_a(k)$ for $k = -L+1 … L-1$ (where $\mathcal{R}(x(n))$ represent the autocorrelation of $x(n)$). Sequences $a(n)$ and $b(n)$ are complementary if their autocorrelation functions add up to a delta pulse (Kronecker or Dirac function): $R_a(k) + R_b(k) = \delta(k)$.

Boolean functions $n_k$ are individual bits of a number $n$ in the binary number system represented by 0 (False) or 1 (True). $n_1$ is the LSB (least significant bit) and $n_N$ is the MSB for $n = 0 … L - 1$, where $L = 2^N$. Boolean functions $n_k$ are periodic functions of $n$ having period $2^k$. $n$ can be represented by Boolean functions as: $n = \sum_{k=1}^{N} n_k \cdot 2^{k-1}$.

Bipolar Walsh functions are defined as: $W_l(n) = \prod_{k=1}^{N}(-1)^{l_k \cdot n_k}$

Elements of a $M \times M$ matrix $\boldsymbol{U}$ are denoted by $\boldsymbol{U}_{r,s}$, where r and s are the row and the column index, respectively. If the matrix is a function of $n$, its elements are $\boldsymbol{U}_{r,s}(n)$.

A wide-sense unitary matrix is defined as: $\boldsymbol{U} \cdot \boldsymbol{U}^H = K \cdot \boldsymbol{I}$ where K is a constant (when $K=1$ it is strict-sense unitary). A general form of a 2 X 2 unitary matrix is: $\boldsymbol{U} = \begin{bmatrix} C & S \\ -S^* & C^* \end{bmatrix}$ where C and S are two complex numbers. We have that $\boldsymbol{U}_{0,0} = C$, $\boldsymbol{U}_{0,1} = S$, $\boldsymbol{U}_{1,0} = -S^*$, $\boldsymbol{U}_{1,1} = C^*$.

Following [16], the complementary sequences are represented by a 2 X 2 matrix $\mathcal{M}(n)$, which contains in addition to a complementary pair $a(n)$ and $b(n)$ their reverted/conjugated sequences ($a^*(n)$ is a complex conjugate version of $a(n)$ and $a^R(n)$ is the time reverted version of $a(n)$ $a^R(n) = a(L - n - 1)$.):

$$\mathcal{M}(n) = \begin{bmatrix} a(n) & b(n) \\ -b^{*R}(n) & a^{*R}(n) \end{bmatrix} \tag{1}$$

Since $a^{*R}(n)$ has the same autocorrelation as $a(n)$, $a^{*R}(n)$ and $b(n)$ also form a complementary pair, as do $a(n)$ and $b^{*R}(n)$. Hence, every row/column of the complementary matrix $\mathcal{M}(n)$ consists of a complementary pair.

# The general standard algorithm

The following universal algorithm generates all standard sequences of length $L = 2^N$. This algorithm cannot generate non-standard complementary sequences and derived sequences. Also this algorithm generates some duplicate sequences in QAM constellations and this issue must be treated separately (see discussion in [5]).

The general form of the algorithm is multiplicative. Factors in the product can be represented in three different ways:

- The *Boolean index form*. The Boolean functions are used to index 2 X 2 unitary matrices.
- The *Boolean exponent form*. Boolean operations (e.g., *nor*, *and*) are used to create new logic functions, which are used as exponents of the unitary matrices element.
- The *algebraic exponent form*. Logic operations are expressed as algebraic operations. Exponents of the unitary matrices element are either first or second order generalized Boolean functions.

The Boolean index form is the most compact form of the proposed generation algorithm. On the other hand, it is the least known form because it does not resemble any previous notation used for representing complementary sequences. However, the next two forms – derived from the Boolean index form - bear much more significant similarity to the known algorithms. Also, they let us derive M-PSK and QAM algorithms as special cases, easily.

## The Boolean index form

Here, $\mathcal{M}_{r,s}^{\{N\}}(n)$ is given by the following scalar product:

$$\mathcal{M}_{r,s}^{\{N\}}(n) = U_{r,n_{P(1)}}^{\{0\}} \cdot U_{B_{P(1)},n_{P(2)}}^{\{1\}} \cdot U_{n_{P(2)},n_{P(3)}}^{\{2\}} \cdot \ldots \cdot U_{n_{P(N-1)},n_{P(N)}}^{\{N-1\}} \cdot U_{n_{P(N)},s}^{\{N\}}$$
$$= U_{r,n_{P(1)}}^{\{0\}} \cdot \left( \prod_{k=1}^{N-1} U_{n_{P(k)},n_{P(k+1)}}^{\{k\}} \right) \cdot U_{n_{P(N)},s}^{\{N\}} \quad (2)$$

where $U^{\{k\}}$ are unitary matrices of the form $U^{\{k\}} = \begin{bmatrix} C_k & S_k \\ -S_k^* & C_k^* \end{bmatrix}$ and $P(.)$ Is a permutation of numbers $1,\ldots,N$.

A more compact form is obtained by introducing functions $\hat{n}_k$ as:

$$\hat{n}_k = \begin{cases} r & \text{for } k = 0 \\ n_{P(k)} & \text{for } k = 1, 2, \ldots, N \\ s & \text{for } k = N + 1 \end{cases}$$

Then we have:

$$\mathcal{M}_{r,s}^{\{N\}}(n) = \prod_{k=0}^{N} U_{\hat{n}_k, \hat{n}_{k+1}}^{\{k\}} \quad (3)$$

The proof of the complementarity of sequences generated by this algorithm is given in Appendix A.

Note that all N+1 factors in the product are scalars. Each scalar is an element of an unitary matrix, which is indexed by a Boolean function. The N+1 unitary matrices can be chosen either in arbitrary manner or to satisfy constellation constraint as discussed in the next section.

## The Boolean exponent form

From (3), we have that:

$$U^{\{k\}}_{\hat{n}_k,\hat{n}_{k+1}} = \begin{cases} U^{\{k\}}_{0,0} = C_k & \text{if } \hat{n}_k = 0 \text{ and } \hat{n}_{k+1} = 0 \\ U^{\{k\}}_{1,1} = C^*_k & \text{if } \hat{n}_k = 1 \text{ and } \hat{n}_{k+1} = 1 \\ U^{\{k\}}_{0,1} = S_k & \text{if } \hat{n}_k = 0 \text{ and } \hat{n}_{k+1} = 1 \\ U^{\{k\}}_{1,0} = -S^*_k & \text{if } \hat{n}_k = 1 \text{ and } \hat{n}_{k+1} = 0 \end{cases}$$

We can also write

$$U^{\{k\}}_{\hat{n}_k,\hat{n}_{k+1}} = (C_k)^{\overline{\hat{n}_k}\&\overline{\hat{n}_{k+1}}} \cdot (C^*_k)^{\hat{n}_k\&\hat{n}_{k+1}} \cdot (S_k)^{\overline{\hat{n}_k}\&\hat{n}_{k+1}} \cdot (-S^*_k)^{\hat{n}_k\&\overline{\hat{n}_{k+1}}} \tag{4}$$

because only one of the exponents is equal to 1 and all others are 0.

Consequently,

$$M^{\{N\}}_{r,s}(n) = \prod_{k=0}^{N}(C_k)^{\overline{\hat{n}_k}\&\overline{\hat{n}_{k+1}}} \cdot (C^*_k)^{\hat{n}_k\&\hat{n}_{k+1}} \cdot (S_k)^{\overline{\hat{n}_k}\&\hat{n}_{k+1}} \cdot (-S^*_k)^{\hat{n}_k\&\overline{\hat{n}_{k+1}}} \tag{5}$$

## The algebraic exponent form

By using the following algebraic expressions to represent logic operations:

$\bar{x} = 1 - x$ and $x \& y = x \cdot y$, we can obtain the algebraic form derived in Appendix B as

$$M^{\{N\}}_{r,s}(n) = C \cdot G^N_{r,s}(n) \cdot \left\{\prod_{k=0}^{N}(|S_k/C_k|)^{-2\hat{n}_k\hat{n}_{k+1}} \cdot (S_k/C_k)^{\hat{n}_{k+1}} \cdot (-S^*_k/C_k)^{\hat{n}_k}\right\} \tag{6}$$

where $C$ is a complex constant and $G^N_{r,s}(n)$ are binary Golay sequences given by:

$$G^N_{r,s}(n) = \prod_{k=0}^{N}(-1)^{\hat{n}_k \cdot \hat{n}_{k+1}} \tag{7}$$

## Special cases

### 16-QAM

To generate 16-QAM from (5) we start with

$$M_{r,s}^{\{N\}}(n) = \left\{ \prod_{k=0}^{N} (C_k)^{\overline{\hat{n}_k \& \hat{n}_{k+1}}} \cdot (C_k^*)^{\hat{n}_k \& \hat{n}_{k+1}} \cdot (S_k)^{\overline{\hat{n}_k \& \hat{n}_{k+1}}} \cdot (-S_k^*)^{\hat{n}_k \& \overline{\hat{n}_{k+1}}} \right\} \qquad (8)$$

and select the unitary matrix elements in the following manner. For only a single $k$, namely $k=K$, we choose $C_K$ and $S_K$ from the 16-QAM constellation. However, $C_K$ must be restricted to one quadrant of 16-QAM to avoid duplication. For all other $k \neq K$, $C_k = 1$ and $S_k = e^{2\pi i \cdot m_k/4}$ $m_k \in \{0, \dots, 3\}$ (both belong to 4-PSK but $C_k$ is restricted to one quadrant). In [5] this algorithm is referred to as a *single QAM-U matrix algorithm*. This special case of the algorithm is universal because it can generate *ANY* constellation provided that $C_K$ and $S_K$ are chosen from that constellation. Even non-rectangular constellations can be generated in this way. To generate a hexagonal constellation, for $k \neq K$ we select $C_k = 1$ and $S_k = e^{2\pi i \cdot m_k/6}$ $m_k \in \{0, \dots, 5\}$. For additional requirements to generate unique sequences refer to [5].

### 64-QAM

64-QAM can also be generated via a *single QAM-U matrix algorithm.* However, not all 64-QAM sequences can be generated by this algorithm [5]. So we need to use two QAM-U matrices at two different locations. All the other matrices use unimodal $C_k = 1$ and $S_k = e^{2\pi i m/4}$. There are two other restrictions on the elements $C_K$ and $S_K$ of the two QAM-U matrices to ensure that such algorithm results in complementary 64-QAM sequences:

- We must use a constellation lattice that is closed under multiplication. For example, if we use the standard lattice that uses odd coordinates for real and imaginary parts of $C_K$ and $S_K$, the product may not be in the lattice. For example, if we use $1 + i$ for both elements, the product in (8) would be: $(1 + i)(1 + i) = 0 + 2i$, as both coordinates are even. Thus, the generated sequence will not belong to the 64-QAM lattice. The solution is to use the natural lattice described in [5], which is closed under multiplication. It is very easy to convert a natural lattice to a standard lattice.
- We must select the values for $C_K$ and $S_K$ that generate sequence elements that lie inside the 64-QAM constellation. For example points $2 + i$ and $1 + 2i$ satisfy this condition because their radius is $\sqrt{5}$. However, points like *3* or $4 + i$ cannot be used because their radii are too big. For detailed treatment of allowed radii see [5].

As it is equivalent to the approach in [5], this algorithm also introduces previously unknown 64-QAM sequences when two QAM-U matrices are used.

## Larger QAM constellation

Larger QAM constellations can use more then two unitary QAM-U matrices. The choice of values of $C$ and $S$ is not as simple as in the case of a one unitary QAM-U matrix. However, the easiest way to determine all acceptable unitary matrices can be determined by computer search.

## M-PSK sequences

We can obtain M-PSK sequences from (6) by using: $C_k = 1, S_k = w_k$, where $|w_k| = 1$, and noting that $w_k^* = w_k^{-1}$. From Appendix C we can see that:

$$\mathcal{M}_{r,s}^{\{N\}}(n) = \widetilde{w}_0 \cdot G_{r,s}^N(n) \cdot \prod_{k=1}^{N} e^{2\pi i \cdot m_k \cdot n_{P(k)}/M} \tag{9}$$

We can call the term $\prod_{k=1}^{N} e^{2\pi i \cdot m_k \cdot n_{P(k)}/M}$ generalized polyphase Walsh function because it reduce to a Walsh function in the binary case $M = 2$. The normalized phase of $\mathcal{M}_{r,s}^{\{N\}}(n)$ is:

$$\sphericalangle\left(\mathcal{M}_{r,s}^{\{N\}}(n)\right)\frac{M}{2\pi} = \frac{M}{2} r \cdot n_{P(1)} + \frac{M}{2} s \cdot n_{P(N)} + m_0 + \frac{M}{2}\sum_{k=1}^{N-1} n_{P(k)} \cdot n_{P(k+1)} + \sum_{k=1}^{N} m_k \cdot n_{P(k)} \tag{10}$$

where $m_k \in \{0 \dots M-1\}$. Here we recognize the algebraic normal form (ANF) from the Davis-Jedwab construction [9,10].

## Binary sequences

From (9) for $M = 2$ we obtain the binary case given by [10]:

$$\mathcal{M}_{r,s}^{\{N\}}(n) = \widetilde{w}_0 \cdot G_{r,s}^N(n) \cdot \prod_{k=1}^{N} e^{(2\pi i \cdot m_k \cdot n_{P(k)})/M} = \widetilde{w}_0 \cdot G_{r,s}^N(n) \cdot \prod_{k=1}^{N}(-1)^{m_k \cdot n_{P(k)}} =$$

$$= \widetilde{w}_0 \cdot G_{r,s}^N(n) \cdot \prod_{k=1}^{N}(-1)^{m_{\hat{P}(k)} \cdot n_k} = \widetilde{w}_0 \cdot G_{r,s}^N(n) \cdot W_l(n) \tag{11}$$

where $l_k = m_{\hat{P}(k)}$ ($l_k$ is a Boolean function of $l$) and $W_l(n)$ is a Walsh function. Where $\hat{P}$ is the inverse permutation of $P$.

## Conclusion

A general form of a generator of complementary sequences is given. It is based on a product of complex variables with exponents that are Boolean polynomials of the second order. Three different forms of this generator are given:

- the Boolean index form,
- the Boolean exponent form,

- the algebraic exponent form

It is shown that the proposed general generator reduces to the well-known generators for binary and M-PSK sequences. However, in the case of QAM sequences the algorithm is novel. It follows the principles of the PU algorithm in [5]. Compared to other algorithms based on Boolean functions, it introduces new 64-QAM sequences when two QAM-U matrices are used. As opposed to usual algorithms that construct 64-QAM sequences as a sum of 4-PSK sequences with 1-2-4 weights, our Boolean index form algorithm is based on a product of scalars that are elements of unitary matrices. The algorithm can also generate 256-QAM, 1024-QAM and larger constellation sequences. Sequences whose elements are in other non-standard constellations, including the hexagonal constellation, can be constructed with the proposed algorithm.

## The program

The algorithm in the Boolean index form implemented as a Matlab code is given in Appendix D. Generating a sequence of length 1024 takes 3ms and generating a sequence of length 64 takes 300 µs (on an average Intel Core2Duo CPU) which is equal to 3 or 2 µs per symbol. If implemented in FPGA it would be easy to achieve generation at speed of a few ns per symbol.

## Appendix A

Our proof is based on mathematic induction. Without loss of generality, we consider only the case $P = [1, 2, 3, \ldots, N]$.

First we verify the result for $N = 0$:

$$\mathcal{M}_{r,s}^{\{1\}}(n) = U_{r,s}^{\{0\}}$$

This is a length 1 sequence, so there are no sidelobes so they are complementary by definition.

Next we prove that if $a_N(n)$ and $b_N(n)$ are complementary for some *N,* they are also complementary for *N+1*.

$$a_{N+1}(n) = \mathcal{M}_{0,0}^{\{N+1\}}(n) = U_{r,n_{P(1)}}^{\{0\}} \cdot \left( \prod_{k=1}^{N} U_{n_k,n_{k+1}}^{\{k\}} \right) \cdot U_{n_{N+1}(n),s}^{\{N+1\}}$$

$$a_{N+1}(n) = U_{r,n_{P(1)}}^{\{0\}} \cdot \left( \prod_{k=1}^{N-1} U_{n_k,n_{k+1}}^{\{k\}} \right) \cdot U_{n_N,n_{N+1}}^{\{N\}} \cdot U_{n_{N+1},s}^{\{N+1\}}$$

for $n < 2^N$: $n_{N+1} = 0$ and for $n \geq 2^N$: $n_{N+1} = 1$ so we have:

$$a_{N+1}(n) = \begin{cases} U_{r,n_1}^{\{0\}} \cdot \left( \prod_{k=1}^{N-1} U_{n_k,n_{k+1}}^{\{k\}} \right) \cdot U_{n_N,0}^{\{N\}} \cdot U_{0,0}^{\{N+1\}} & \text{for } n < 2^N \\ U_{r,n_1}^{\{0\}} \cdot \left( \prod_{k=1}^{N-1} U_{n_k,n_{k+1}}^{\{k\}} \right) \cdot U_{n_N,1}^{\{N\}} \cdot U_{1,0}^{\{N+1\}} & \text{for } n \geq 2^N \end{cases}$$

$$a_{N+1}(n) = \begin{cases} a_N(n) \cdot \boldsymbol{U}_{0,0}^{\{N+1\}} & \text{for } n < 2^N \\ b_N(n) \cdot \boldsymbol{U}_{1,0}^{\{N+1\}} & \text{for } n \geq 2^N \end{cases}$$

Also we have:

$$b_{N+1}(n) = \begin{cases} a_N(n) \cdot \boldsymbol{U}_{0,1}^{\{N+1\}} & \text{for } n < 2^N \\ b_N(n) \cdot \boldsymbol{U}_{1,1}^{\{N+1\}} & \text{for } n \geq 2^N \end{cases}$$

Finally we can write:

$$a_{N+1}(n) = \left[\left(a_N(n) \cdot \boldsymbol{U}_{0,0}^{\{N+1\}}\right), \left(b_N(n) \cdot \boldsymbol{U}_{1,0}^{\{N+1\}}\right)\right]$$

$$b_{N+1}(n) = \left[\left(a_N(n) \cdot \boldsymbol{U}_{0,1}^{\{N+1\}}\right), \left(b_N(n) \cdot \boldsymbol{U}_{1,1}^{\{N+1\}}\right)\right]$$

Where $[x, y]$ represents concatenation. As $\boldsymbol{U}_{N+1}$ is a unitary matrix if $a_N(n)$ and $b_N(n)$ form a complementary pair so do $a_{N+1}(n)$ and $b_{N+1}(n)$.

QED

# Appendix B

By using the following algebraic expressions to represent logic operations:

$\bar{x} = 1 - x$ and $x \& y = x \cdot y$ we can obtain the algebraic formulation:

$$\boldsymbol{M}_{r,s}^{\{N\}}(n) = \left\{ \prod_{k=0}^{N} (C_k)^{(1-\hat{n}_k) \cdot (1-\hat{n}_{k+1})} \cdot (C_k^*)^{\hat{n}_k \cdot \hat{n}_{k+1}} \cdot (S_k)^{(1-\hat{n}_k) \cdot \hat{n}_{k+1}} \cdot (-S_k^*)^{\hat{n}_k \cdot (1-\hat{n}_{k+1})} \right\} \quad (12)$$

$$= \left\{ \prod_{k=0}^{N} (C_k)^{1-\hat{n}_k-\hat{n}_{k+1}+\hat{n}_k \hat{n}_{k+1}} \cdot (C_k^*)^{\hat{n}_k \cdot \hat{n}_{k+1}} \cdot (S_k)^{\hat{n}_{k+1}-\hat{n}_k \hat{n}_{k+1}} \cdot (-S_k^*)^{\hat{n}_k - \hat{n}_k \hat{n}_{k+1}} \right\}$$

$$= \left\{ \prod_{k=0}^{N} C_k \cdot (-S_k S_k^* / C_k C_k^*)^{-\hat{n}_k \hat{n}_{k+1}} \cdot (S_k / C_k)^{\hat{n}_{k+1}} \cdot (-S_k^* / C_k)^{\hat{n}_k} \right\}$$

$$= \prod_{k=0}^{N} C_k \cdot \prod_{k=0}^{N} (-1)^{\hat{n}_k \hat{n}_{k+1}} \cdot \prod_{k=0}^{N} (|S_k / C_k|)^{-2\hat{n}_k \hat{n}_{k+1}} \cdot (S_k / C_k)^{\hat{n}_{k+1}} \cdot (-S_k^* / C_k)^{\hat{n}_k}$$

$$= C \cdot \boldsymbol{G}_{r,s}^{N}(n) \cdot \left\{ \prod_{k=0}^{N} (|S_k / C_k|)^{-2\hat{n}_k \hat{n}_{k+1}} \cdot (S_k / C_k)^{\hat{n}_{k+1}} \cdot (-S_k^* / C_k)^{\hat{n}_k} \right\} \quad (13)$$

where $C = \prod_{k=0}^{N} C_k$ and

$$\boldsymbol{G}_{r,s}^{N}(n) = \prod_{k=0}^{N} (-1)^{\hat{n}_k \cdot \hat{n}_{k+1}} = (-1)^{r \cdot n_{P(1)}} \cdot \left( \prod_{k=1}^{N-1} (-1)^{n_{P(k)} \cdot n_{P(k+1)}} \right) \cdot (-1)^{n_{P(N)} \cdot s}$$

The phase of $G_{r,s}^N(n)$ is

$$g_{r,s}^N(n) = \pi\left(\sum_{k=0}^{N} \hat{n}_k \cdot \hat{n}_{k+1}\right) = \pi\left(r \cdot n_{P(1)} + s \cdot n_{P(N)} + \sum_{k=1}^{N-1} n_{P(k)} \cdot n_{P(k+1)}\right) \tag{14}$$

(12) generates two binary Golay sequences and its reverses. So we have:

$$a_N(n) = G_{0,0}^N(n) = \left(\prod_{k=1}^{N-1}(-1)^{n_{P(k)} \cdot n_{P(k+1)}}\right)$$

$$b_N(n) = G_{1,0}^N(n) = (-1)^{n_{P(1)}} \cdot G_{0,0}^N(n) = (-1)^{n_{P(1)}} \cdot a_N(n),$$

$$b_N^R(n) = G_{0,1}^N(n) = (-1)^{n_{P(N)}} \cdot G_{0,0}^N(n) = (-1)^{n_{P(N)}} \cdot a_N(n),$$

$$b_N^R(n) = G_{0,1}^N(n) = (-1)^{n_{P(1)}} \cdot (-1)^{n_{P(N)}} \cdot G_{0,0}^N(n) = (-1)^{n_{P(1)}} \cdot (-1)^{n_{P(N)}} \cdot a_N(n),$$

We can also write: $a_N(n) \cdot b_N(n) = (-1)^{n_{P(1)}}$ which is a simple test if a complementary pair is standard or non-standard. Also we have: $a_N(n) \cdot b_N^R(n) = (-1)^{n_{P(N)}}$ because they are also complementary. As $b_N^R(n)$ is easily determined from $b_N(n)$, $P(1)$ and $P(N)$ are easily determined for each complementary pair.

## Appendix C

Starting from (6) and inserting: $C_k = 1, S_k = w_k$, where $|w_k| = 1$, and noting that $w_k^* = w_k^{-1}$ we get

$$\mathcal{M}_{r,s}^{\{N\}}(n) = G_{r,s}^N(n) \cdot \left\{\prod_{k=0}^{N} w_k^{\hat{n}_{k+1}} \cdot \left(-w_k^{-\hat{n}_k}\right)\right\} = G_{r,s}^N(n) \cdot \left\{\prod_{k=0}^{N} w_k^{\hat{n}_k} \cdot \left(-w_k^{-\hat{n}_{k+1}}\right)\right\} =$$

$$= G_{r,s}^N(n)\left\{w_N^s \prod_{k=0}^{N-1} w_k^{n_{P(k)}}\right\} \cdot \left\{w_0^{-r} \prod_{k=1}^{N}\left(-w_k^{-n_{P(k)}}\right)\right\} = G_{r,s}^N(n)\left\{w_N^S \prod_{k=1}^{N} w_{k-1}^{n_{P(k)}}\right\} \cdot \left\{w_0^{-R} \prod_{k=1}^{N}\left(-w_k^{-n_{P(k)}}\right)\right\}$$

$$= (w_N^S w_0^{-R}) \cdot G_{r,s}^N(n) \cdot \prod_{k=1}^{N} w_{k-1}^{n_{P(k)}} \cdot -w_k^{-n_{P(k)}} = (w_N^S w_0^{-R}) \cdot G_{r,s}^N(n) \cdot \prod_{k=1}^{N}\left(-\frac{w_{k-1}}{w_k}\right)^{n_{P(k)}}$$

Let $\tilde{w}_0 = (w_N^S w_0^{-R})$ and $\tilde{w}_k = -w_{k-1}/w_k$, which are also unimodular. Now we have:

$$\mathcal{M}_{r,s}^{\{N\}}(n) = \tilde{w}_0 \cdot G_{r,s}^N(n) \cdot \prod_{k=1}^{N} \tilde{w}_k^{n_{P(k)}}$$

In the case of M-PSK $\tilde{w}_k = e^{2\pi i \cdot m_k/M}$ where $m_k \in \{0 \dots M-1\}$ and k = 0 … N

$$\mathcal{M}_{r,s}^{\{N\}}(n) = \tilde{w}_0 \cdot G_{r,s}^N(n) \cdot \prod_{k=1}^{N} e^{2\pi i \cdot m_k \cdot n_{P(k)}/M}$$

We can call the term $\prod_{k=1}^{N} e^{2\pi i \cdot m_k \cdot n_{P(k)}/M}$ generalized polyphase Walsh function because it reduce to a Walsh function in the binary case $M = 2$.

Taking only the phase terms (with $g_{r,s}^N(n)$ the phase of the binary Golay sequences given in Appendix B) we have:

$$\sphericalangle\left(\mathcal{M}_{r,s}^{\{N\}}(n)\right) = g_{r,s}^N(n) + 2\pi/M\, m_0 + 2\pi/M \cdot \sum_{k=1}^{N} m_k \cdot n_{P(k)}$$

$$= \pi \cdot r \cdot n_{P(1)} + \pi \cdot s \cdot n_{P(N)} + 2\pi/M\, m_0 + \pi \cdot \sum_{k=1}^{N-1} n_{P(k)} \cdot n_{P(k+1)} + 2\pi/M \cdot \sum_{k=1}^{N} m_k \cdot n_{P(k)}$$

$$= 2\pi/M \cdot \left(\frac{M}{2} r \cdot n_{P(1)} + \frac{M}{2} s \cdot n_{P(N)} + m_0 + \frac{M}{2} \sum_{k=1}^{N-1} n_{P(k)} \cdot n_{P(k+1)} + \sum_{k=1}^{N} m_k \cdot n_{P(k)}\right)$$

## Appendix D

```
function X=booleanCS(P,C,S,r,s);
% P(k) is a permutation of k=1:N
% C(k), S(k) complex numbers for k=1:N+1
% depending on (r, s) (that can have values: 0 or 1) X is one of the sequences:
% (0,0): "a", (0,1): "b", (1,0): "b reverted/conjugated/negated" and (1,1): "a reverted/conjugated"
N=length(P); L=2^N;                                 % Number of iterations N,  sequence length L
CS=[C;S;-conj(S);conj(C)];                          % Elements of the QAM-U matrix
b=(0.5*ones(1,N)).^(0:N-1); b=b'*(0:L-1);b=floor(rem(b,2));   % Calculating Boolean functions n_k
B=[r*ones(1,L);b(P,:);s*ones(1,L)];                 % Extended Boolean functions n_k hat
X=1; for k=1:N+1; X=X.*CS(1+B(k,:)+2*B(k+1,:),k); end         % Calculating a sequence
```